\documentclass[12pt,preprint]{aastex}

\usepackage{amsmath}

\slugcomment{Not to appear in Nonlearned J., 45.}

\shorttitle{Pakal}
\shortauthors{De la Luz et al.}

\begin{document}

\title{Pakal:A 3D Model to Solve the Radiative Transfer Equation}

\author{Victor De la Luz\altaffilmark{1}, Alejandro Lara\altaffilmark{1}, J.E. Mendoza-Torres\altaffilmark{2} and Caius L. Selhorst\altaffilmark{3,4}}
\altaffiltext{1}{Instituto de Geof\'isica, Universidad Nacional Aut\'onoma de M\'exico, M\'exico, 04510.}
\altaffiltext{2}{Coordinaci\'on de Astrof\'isica, Instituto Nacional de Astrof\'isica \'Optica y Electr\'onica, M\'exico, 72840.}
\altaffiltext{3}{CRAAM, Universidade Presbiteriana Mackenzie, S\~ao Paulo, SP, Brasil, 01302-907.}
\altaffiltext{4}{IP\&D, Universidade do Vale do Para\'iba, S\~ao Jos\'e dos Campos, SP, Brasil, 01302-907.}


\begin{abstract}
We present a new numerical model called ``PAKAL'' intended to solve the radiative 
transfer equation in a three dimensional (3D) geometry, using the approximation
for a locally plane parallel atmosphere.
Pakal uses pre-calculated radial profiles of density and temperature
(based on hydrostatic, hydrodynamic or MHD models) to compute the emission from 3D source structures with 
high spacial resolution. Then, Pakal solves 
the radiative transfer equation  in a set of (3D) ray-paths,
going from the source to the observer. Pakal uses a new algorithm to compute the radiative transfer
equation by using an Intelligent System consisting of three structures: a cellular automaton; 
 an expert system; and a program coordinator.
The code  outputs can be either two dimensional maps or one dimensional profiles, 
which reproduce the observations with high accuracy, 
giving in this way,
detailed physical information about the environment where the radiation was 
generated and/or transmitted.
We present the model applied to
a 3D solar radial geometry, assuming a locally plane-parallel atmosphere, 
and
thermal free-free  
radio emission from a Hydrogen-Helium gas  in 
thermodynamic equilibrium. We also present the convergences test of the code.
We computed the synthetic spectrum of the centimetric - millimetric solar 
emission and found  better agreement with observations (up to $10^4$ K at
20 GHz) than previous models reported in literature. 
%
The stability and convergence test show the high accuracy of the code.
Finally, Pakal can improve the integration time  by up to an order of magnitude 
compared against linear integration codes.
\end{abstract}

\keywords{radiative transfer, solar radio emission, numerical models}

\section{Introduction}
The
observation  and study of 
radio emissions coming from distant sources is a valuable tool to investigate
these objects and the medium between them and the observer.
For instance, by assuming an emission mechanism, we are able to obtain detailed
physical properties of the observed object as the density, temperature, 
magnetic field, etc.

Generally, due to observational limitations, we obtain two
dimensional projections in the plane of the sky of the emission and/or 
absorption in the ray paths of each point of the
observed region inside  the telescope field of view. 
These maps represent not only the emitting object, but all the possible 
flux changes due to the
emission and/or absorption that may take place in the medium between 
the source and the observer. Therefore to get reliable information 
through the study of radio emissions, 
 it is necessary to take into 
account the detailed three dimensional (3D) structure of both, the source and the medium.
In this work we present ``Pakal'', a numerical 
model intended to solve the radiative transfer
equation, designed to study astronomical objects specially in the millimeter
and submillimeter wavelengths.

Now a days, there are few dozens of codes to solve the radiative transfer equation, though, each code
is designed to solve a very specific problem. As instance, there are codes to
solve the radiative transfer equation in  Earth-like 
atmospheres  
\citep[e. g.][]{2006EOSTr..87...52O,2005BAMS...86.1275C,2001AGUSM...U61A01D}. 
More specifically, 
the I3RC Monte Carlo community model of 3D radiative transfer \citep{2005BAMS...86.1275C}; 
the ARTS package \citep{2005JQSRT..91...65B}; 
Battaglia-Mantovani model \citep{2005JQSRT..95..285B}; 
GRIMALDI \citep{2001JGR...10614301S};
MCARaTS \citep{2006JAtS...63.2324I}; 
SHDOM \citep{1998JAtS...55..429E};  
and SHARM-3D \citep{2002ApOpt..41.5607L},
are  designed to study the dispersion of tele-communication radio waves
in the Earth atmosphere.
These models simulate layers of a plane-parallel atmosphere and
are based mainly in Monte Carlo techniques. 

In the astrophysics community, there are mainly  two branches of codes to simulate
the emission of stellar atmospheres: 
\begin{itemize}
\item Codes to simulate the atmosphere structure 
(the variation with height of physical parameters as density, temperature, etc).
\item Codes to compute the synthetic spectrum. 
\end{itemize}

Commonly, the codes for stellar atmospheres simulation deal with  
specific physical 
conditions. For example, ATLAS12 \citep{1979ApJS...40....1K}; MARCS
\citep{1975A&A....42..407G}; and
PHOENIX \citep{1999ApJ...512..377H}
are general propose codes for stellar atmospheres which
take into account only the emissions from the stellar
photospheres.
The PANDORA \citep{1976ApJS...30....1V} and MULTI \citep{1992ASPC...26..499C}
 codes
simulate stellar atmospheres using conditions similar to the solar atmosphere 
but only in the region
below the corona. 
Whereas CHANTI \citep{1997A&AS..125..149D}
 simulates atmospheres
with coronal conditions.
The chromospheric and coronal codes are oriented to reproduce the 
ultraviolet (UV) 
and Visible spectrum, and therefore fail to reproduce observations
 in the radio range 
\citep[]{1991ApJ...370..779Z,1993ApJ...403..426E,2005A&A...433..365S}.

Examples of  codes in the second branch (synthetic radiative spectrum) are:
SYNTHE \citep{1979ApJS...40....1K}; SPECTRUM \citep{1995ApJ...439..875H};
and FANTOM \citep{1991A&A...247..108C}. These codes are complementary to
codes in the former branch and are necessary to compute
the final stellar spectrum. 
We note that PANDORA and CHANTI, from the first group,  are also able to 
compute the spectrum.

Some codes are intended to compute particular stellar atmospheres 
 e. g.:
STERNE3 \citep{2006BaltA..15..115B} for Hydrogen-Deficient Stars; 
LINE-BY-LINE METHOD \citep{2004A&A...428..993S} for stars in early and 
intermediate stage; 
PRO2 \citep{1986A&A...161..177W} and 
TLUSTY \citep{1995ApJ...439..875H} for hot stars;
WM-basic \citep{2001A&A...375..161P} for expanding atmospheres;
CMFGEN \citep{1998ApJ...496..407H} for Wolf-Rayet stars; and
FASTWIND \citep{1997A&A...323..488S} for stars with high mass loss.

Pakal is a completely new code which can be applied to any 
geometry, radiation and absorption mechanisms (focused in this work to 
millimeter and submillimeter wavelengths, but easily configurable for
other wavelengths).
This flexibility is achieved by the means of four completely independent 
modules:
the numerical model (Section \ref{rtt}); geometry model (Section \ref{sec:gepmetry}); numerical methods; and physical functions (Section \ref{sec:physics}).

Pakal uses a new method to compute the radiative transfer equation in a 
set of 3D ray paths,
this is an
intelligent system called ``Tulum'' 
(Section \ref{sec:tulum}) which helps
to reduce the integration time up to one order of magnitude as compared against
 direct
integration codes (Section \ref{sec:results}).
%
Pakal is able to compute the contributions to the opacity function of each 
chemical element and its ionization states. To accomplish this, the code needs   as input,
detailed profiles of electron temperature and ion densities (Appendix \ref{sec:testing}).

\section{Radiative Transfer Theory}\label{rtt}
\begin{figure}
\begin{center}
\includegraphics[width=1.0\textwidth]{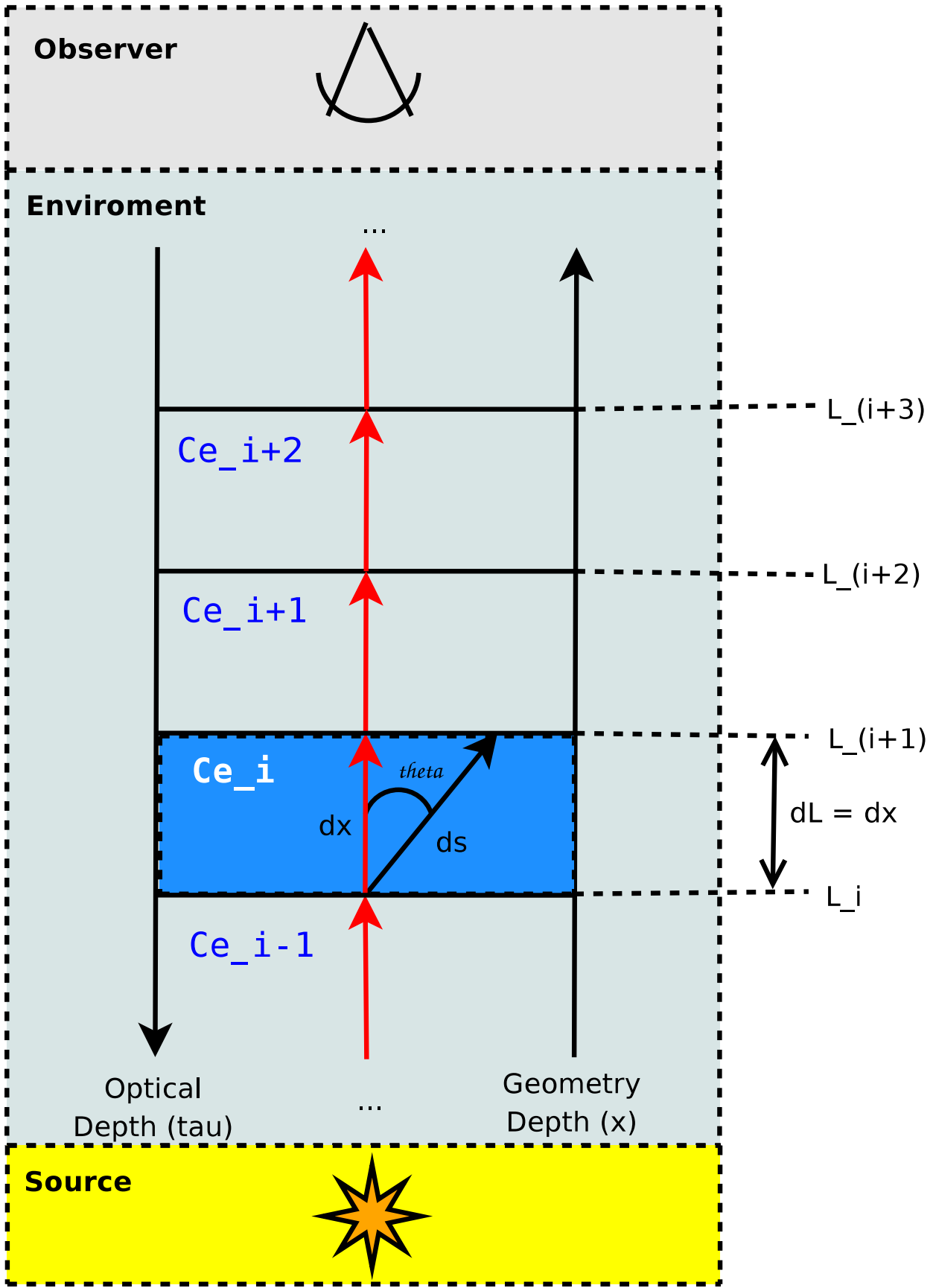}
\caption{General geometry used in the numerical model.}\label{modelonumerico}
\end{center}
\end{figure}

The specific intensity is the most basic entity in radiative transfer theory, 
and is defined as the amount of energy $dE$ passing though an area $dA$, during
a time $dt$, coming inside a solid angle $dw$, in an interval of frequency
$d\nu$, with a direction given by $\hat{r}$ \citep{1986tra..book.....R}
$$dI_{\nu} = \frac{dE}{dA~dt~dw~d\nu~\hat{r}\cdot\hat{n}} ,$$
where  $\hat{r}$ and $\hat{n}$ are the direction and normal (to $dA$)  
unitary  vectors,
respectively and can be written as,
$$\hat{r}\cdot\hat{n} = \cos\theta = \mu ,$$
where $\theta$ is the angle between  $\hat{r}$ and $\hat{n}$.
When  radiation interacts with matter, crossing a distance $ds$, 
the change in the
specific intensity $dI_{\nu}$
is equal to the emission of the medium, $\epsilon_{\nu}$, minus the
radiative energy absorbed by the medium, $\kappa_{\nu}I_{\nu}$, this is \citep{1960ratr.book.....C}: 
\begin{equation}\label{et1}
\frac{dI_{\nu}}{ds}=-\kappa_{\nu}I_{\nu}+\epsilon_{\nu},
\end{equation}
where $\kappa_{\nu}$ is the opacity function which depends on the physical
properties of the medium.
Assuming a plane-parallel atmosphere, it is possible to write $ds$,
in terms of the geometric distance $dx$ (see Figure \ref{modelonumerico}),
$$dx = ds\cos(\theta) = \mu ds;$$
then, using the optical depth, 
$d\tau_{\nu} = - \kappa_{\nu}dx$; and Kirchhoff's law 
($\epsilon_{\nu} = \kappa_{\nu}S_{\nu}$), Eq. \ref{et1} may be written as:
$$\frac{dI_{\nu}}{d\tau_{\nu}}-\frac{I_{\nu}}{\mu}=-\frac{S_{\nu}}{\mu}.$$
The solution, in the  [${\tau_{1,\nu}}$, ${\tau_{2,\nu}}$]  optical depth interval 
(where ${\tau_{1,\nu}} > {\tau_{2,\nu}}$) is
\begin{eqnarray}\label{egtr}
I_{\nu}(\tau_{2,\nu} )& =& I_{\nu}(\tau_{1,\nu})e^{-(\tau_{1,\nu}-\tau_{2,\nu})/\mu }\nonumber\\
& & - \frac{1}{\mu}\int_{\tau_{1,\nu}}^{\tau_{2,\nu}}S_{\nu}(\tau_{\nu})e^{-(\tau_{\nu}-\tau_{2,\nu})/\mu}d\tau_{\nu}
\end{eqnarray}

For solar conditions, the scattering is negligible in the millimeter and
 submillimeter wavelength range  \citep{1976ApJS...30....1V}. 
Therefore for our purposes Eq. \ref{egtr} is completely valid.

Assuming $\tau_2 < \tau_1$; $\mu = 1$; and a source function 
constant in each cell ``$i$'' ($0\leq i \leq n$), we integrate 
Eq. \ref{egtr} in
an array of $n$ consecutive cells  (see Figure \ref{modelonumerico}), using: 
\begin{eqnarray}\label{elmodelonumerico}
I_{\nu}(L_{i+1}) = I_{\nu}(L_i)\exp\left[-\frac{dL}{2}(k_{\nu}(L_i) + k_{\nu}(L_{i+1}))\right]\nonumber \\
 + S_{\nu}(L_i+0.5dL)\left(1 - \exp\left[-\frac{dL}{2}(k_{\nu}(L_i) + k_{\nu}(L_{i+1}))\right]\right).
\end{eqnarray}
where $I(L_i)$ is the specific intensity (coming) from the cell ``$i-1$''; 
$I(L_{i+1})$ is the specific intensity (getting out) of cell ``$i$''; and $dL$ is the
integration step. The computation of each $L_i$ is done by the geometry module (see Section \ref{sec:gepmetry}).

\section{Pakal Model}
Pakal (the name of the king of Palenque in the Mayan Culture) is written in C language, using an object oriented technique 
\citep{cavanzado}
which allow us to encapsulate sets of common
properties or functions in libraries.
The code is based on four independent modules: 
i) the numerical model, ii) the geometry, 
iii) numerical methods and iv) physical functions.

Once the geometry is defined, Pakal generates a series of independent
ray paths, from the source to the observer, reads 
pre-defined  temperature and density profiles  and, if necessary, performs 
an interpolation of the readed values, covering a larger number of points 
in altitude.
Then, using 
an intelligent system called Tulum 
solves 
the radiative transfer equation 
(the related algorithms are part of the numerical module). 

\subsection{Tulum: The Intelligent System} \label{sec:tulum}
\begin{figure}
\begin{center}
\includegraphics[width=1.0\textwidth]{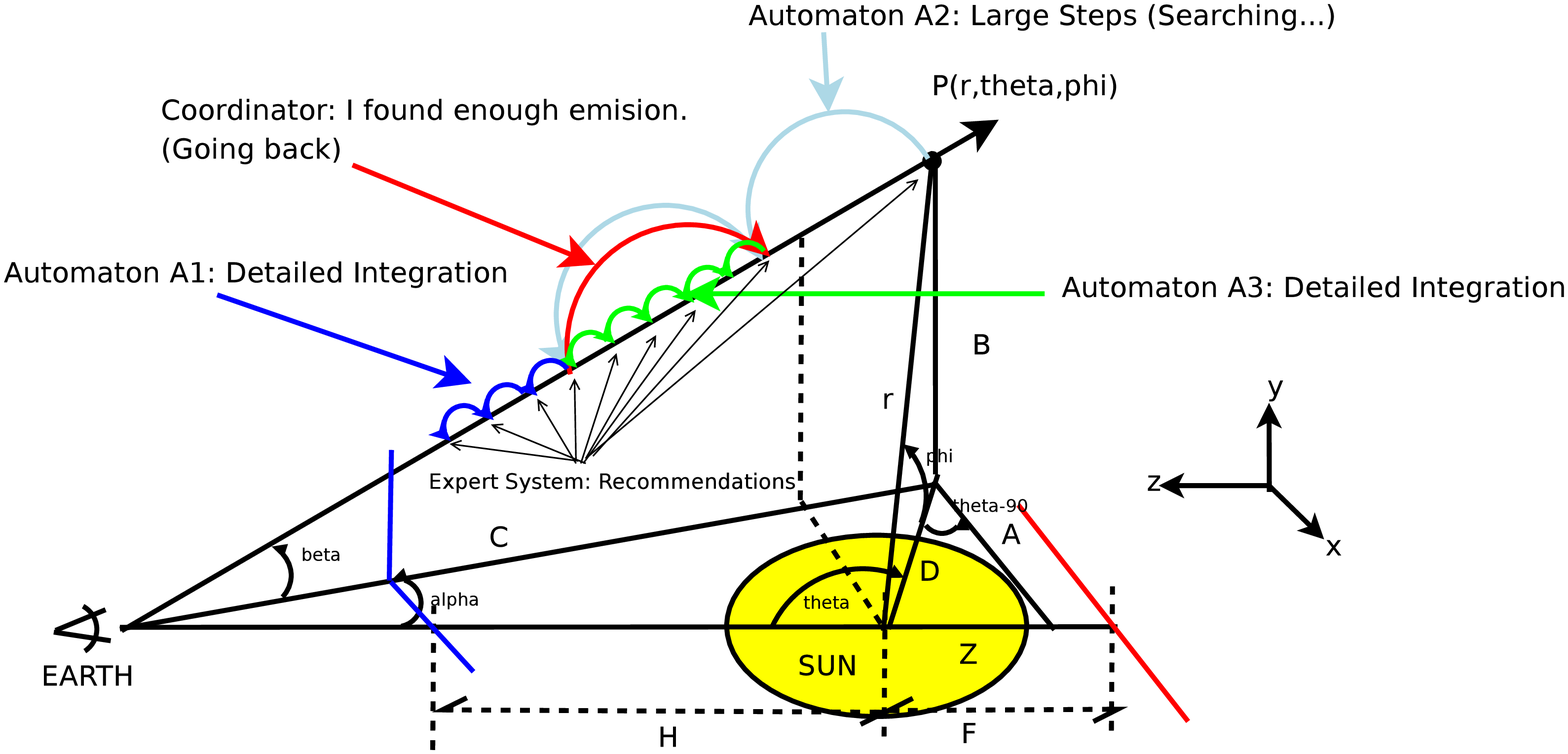}
\caption{Detailed geometry used in Pakal showing, in an (Earth - P) ray path, 
three possible states of the integration process. 
}\label{geometriadetulum}
\end{center}
\end{figure}

The Intelligent System used in Pakal is called Tulum and helps to 
solve the radiative transfer equation in a new and 
very efficient way. In Figure \ref{geometriadetulum}  a schematic
diagram of the automaton is presented. 

Tulum is formed by three independent components: 
\begin{itemize}
\item A coordinator which controls each step of the integration process.
The coordinator uses the
recommendations of the expert system and the states of the cellular automaton to 
decide the next stage of the integration process.
\item An expert system who recommends, based on the current status (position 
and physical conditions), whether or not it is necessary to 
integrate in this point and, if necessary, recommends a change of the
integration step size.
\item A cellular automaton,  with a set of previously established states, 
which is able to save the current status of the integration process.
\end{itemize}

Tulum can integrate numerically any given function 
(not only the radiative transfer equation). 
The integration process  is carried out in the following way:

\begin{enumerate}
\item When the coordinator program receives the spatial coordinates of
two contiguous integration points (from the 3D geometry module),
he looks for the physical conditions
at these points (from the temperature and density radial models). 
If necessary, the numerical module of Pakal automatically interpolates 
the radial  temperature and density models,
at the specific points, using either of two classical methods: 
linear or cubic spline interpolations. In this work we use linear 
interpolation
(the cubic spline interpolation fails because the temperature profile has a very
large gradient in the Solar Transition Zone).

Once the coordinator knows the physical conditions of the medium, he asks for
a recommendation to the expert system,
and also asks for the present state of the cellular
automaton. 
Based on this information,
the coordinator can take the decision of either going 
ahead with the integration 
process (using small or large steps); or going backwards.
Then, the coordinator computes the emission 
(using the numerical module, see  section \ref{sec:physics}); 
and updates the current 
state of the automaton via the $\epsilon$ variable (which
is used to switch between two automaton states).


The set of possible decisions (as shown in Table \ref{tabladecision}), are based on two considerations:

\begin{itemize}
\item In order to save computation time, we neglect the emission that does not contribute to the total 
brightness temperature.
\item On the other hand, we include, with a high spatial resolution, the 
emission of any structure in the solar atmosphere, which contribute to the total brightness temperature. 
\end{itemize}


\item The second component is the expert system which, based on the
physical conditions of each specific point,  decides 
if it is useful to integrate on this region 
and recommends the size of the following integration step.
The recommendation is based on two plasma parameters: the 
plasma frequency ($\nu_p=9\times 10^{-3}\sqrt{n_e}\mbox{ [MHz]}$) and the 
minimal emission. 

The plasma frequency
is important to obtain  the position (atmospheric height)
of the interface between regions where electromagnetic waves, at any given 
frequency, can propagate or not. 

The minimal emission parameter defines the lower limit where the 
local emission is negligible and also
controls both, the error due to this neglected flux; and 
the performance of the integration
process, saving in this way a large amount of computation time.
There is also another numerical error, associated to the small and large steps.
We present the analysis of convergence of these errors in Appendix \ref{apend:conv}.
The minimal emission can be set by the user 
via ``-min'' parameter 
at the console or can be managed 
automatically by the code (see section \ref{sec:minparameter}). 
%




The expert system can recommends the integration steps (small or large), based on the following cases:
\begin{itemize}
\item If $\nu_p > \nu$, then the wave can not propagate and the
 experts recommends a small integration step. 
We consider this case, because we want to know the height where the radiation 
starts propagating in the atmosphere.

\item If $\nu_p \leq \nu$  and the local emission 
is greater than the minimal emission (the amount of emission is important).
The wave can propagate and
the expert recommends small integration step
(we want to analyze in detail the emission process).

\item If $\nu_p \leq \nu$ and the local emission is lower than the minimal emission, the wave can propagate but there is not enough emission, therefore  the expert recommends a large integration step 
(we want to save time in the computation process).
\end{itemize}

The recommendations are managed by two variables: ``q'', the local behavior  
of the emission
and ''y'',  the size of the integration step  
(see Table \ref{tablayq}). These variables are transmitted to the 
coordinator,
as well as
the variable ``state'' which contains the current state
of the Cellular Automaton (Table \ref{tabladecision}).


\item The Cellular Automaton is the logical structure that stores the 
 stage
of the integration process, is the memory of the system. It
is controlled by two parameters: $\epsilon$ (the variable that preserves
the memory when the system switch from one state to another); and the stack, 
 a logical structure which preserves a local memory 
of the automaton states. The stack help us to control the number of steps
when the coordinator decides to ``go back'' in the integration process,
and is represented by two integer variables: ``i'' and ``n'' 
(where n=largestep/smallstep, 
in this way, we warranty that the total length of the small and large
steps is the same when the system enter in the ``go back'' process).

The automaton can be in any of the four following states (see Table \ref{tablaestado}):
\begin{itemize}
\item A1: Integrating using small steps.
\item A2: Integrating using large steps.
\item A3: ``Going back'': I tried to integrate using large steps
 but I had to return because the local emission is larger than the 
minimal emission. Therefore I will
integrate with small steps up to the returning point (this state 
shows the necessity of the Stack structure). 
\item A4: Something is wrong. This is an error.
\end{itemize}
\end{enumerate}

In summary, the three components of Tulum conform a very efficient
intelligent system to 
switch between integration steps; control the associated errors;
and reproduce the emission with high spatial resolution.

\subsection{3D Geometrical Model}\label{sec:gepmetry}
The geometrical model
was designed to optimize the computations of  solar 3D structures based on 
radial profiles of physical parameters (in general, quiet Sun models for the electronic density and temperature 
are given as radial profiles, starting at photospheric level
and extending to different atmospheric altitudes). 

The origin of the coordinated system is located
 at the center of the solar sphere,
the Z axis points towards the observer, the Y axis points to the solar
North, and the X axis completes the system.
In this geometry, a ray path 
from a given point in the plane of the sky 
to the observer 
is formed by a set of radial vectors (see Figure \ref{geometriadetulum}).
 These vectors describe both, the
integration mesh and the radial values of density and temperature
along this ray path. 
The radiative transfer equation is integrated along each ray path and 
the set of 
ray paths forms the 2D projection (on the plane XY) of the 3D model.

In this geometry each point of the mesh is defined as:
$$\vec{r}_{\alpha_x,\beta_y}(z) = (r(\alpha_x,\beta_y,z),\theta(\alpha_x,\beta_y,z), \phi(\alpha_x,\beta_y,z))$$
where $r$ is the module of vector $\vec{r}$ ; $\theta$ the angle between the Z axis and 
the projection of  $\vec{r}$ on the XZ plane; $\phi$ is
the angle between  $\vec{r}$ and the ZY plane and $z$ is
the projection of  $\vec{r}$ on to the Z axis.
From the observer point of view, each ray path represents
a pixel $(x,y)$ on the projected 2D image and is defined by the angles 
$\alpha_x$ and $\beta_y$.
Each ray path is divided in $k$ points separated by a distance $dl$, 
for simplicity, we do not use directly  $dl$ but its projection, $dz$,
on the Z axis.

The mesh is defined by two constants:  the Astronomical Unit, UA = $1.5\times 10^8$ km and the solar radii  $R_{\odot}= 6.96\times 10^5$ km; plus
the following variables:
\begin{itemize}
\item $n$: The image resolution is $n \times n$  pixels.
\item $x$: Variable in the X direction ranging from $-(n-1)/2$ to $(n-1)/2$.
\item $y$: Variable in the Y direction ranging from $-(n-1)/2$ to $(n-1)/2$.
\item $R_T$: The maximum radial distance, in the 2D projection, considered for the integration (we use  $R_T = 2R_{\odot}$).
\item $F$: Defines (in units of solar radii), the starting point of the 
integration process,
$F=0$ means that the starting point lays in the plane XY; $F=-1$ 
in a parallel plane 
located at one solar radii behind of the origin; and $F=1$ in a parallel plane 
located at one solar radii in front of the origin (by default, we use
 $F= -R_T$).
\item $H$: is the final point of the integration process, the default 
value is  $H= R_T$.
\item $dl$: integration step in km.
\end{itemize}

The process starts by computing the matrix of angles:
$$M_{n,n} = \left\{ (\alpha(x),\beta(y))\ |\ \frac{-(n-1)}{2} \leq x,y \leq \frac{n-1}{2}\right\}$$
Then, the initial and final integration points are computed 
for each element  $M_{x,y}$. These points are defined by the user ($F$ and $H$,
respectively). It is possible that some ray paths intersect the
solar surface, for such cases we define $F = z_0,$ where  $z_0$
is the projection of the intersection point
 on the Z axis.
Once the initial and final integration points are known the code generates
 the set of points:
$$L_{x,y} = \{r(\alpha_x,\beta_y,z),\theta(\alpha_x,\beta_y,z), \phi(\alpha_x,\beta_y,z)\ |\ z_0 \leq z \leq H \mbox{ and } z = m*dz, m \in N \},$$
and solve Eq. \ref{elmodelonumerico}.

\subsection{Model for thermal radio emission}\label{sec:physics}
At quiet regions, the main contribution to the emission and absorption
is due to  free - free interactions, in particular, 
free electrons interacting with ions. The electron - electron; ion - ion; and free - bound interactions, do not
contribute significantly to the total emission \citep{1985ARA&A..23..169D}.
Even more, for radio emissions, only distant  electron - ion interactions  are 
important \citep{1985ARA&A..23..169D}.
Therefore, in this case, the absorption coefficient is \citep{1985ARA&A..23..169D}:
\begin{equation}\label{fullopa}
\kappa_{\nu} = \sum_i\frac{1}{3c}\left(\frac{2}{\pi}\right)^{1/2}\frac{\nu_p^2}{\nu^2}\frac{4\pi Z_i^2n_ie^4}{m^{1/2}(kT^{3/2})}\frac{\pi}{\sqrt{3}}G(T,\nu),
\end{equation}
where $n_i$ is computed using Saha equation \citep{1961psc..book.....A}:
\begin{equation}\label{saha}
\log\frac{n_{i+1}}{n_i} = -0.1761-\log(P_e)+\log\frac{u_{i+1}}{u_i}+2.5\log T -\chi_{i}\frac{5040}{T},
\end{equation}
where
$u_i$ is the statistical weight; $\chi_i$ is the ionization energy;   
$P_e=n_eKT$; and 
$n_e$ is the observed electronic density profile.

Equation \ref{fullopa}
may be approximated, according to the appropriate Gaunt factor to:
\begin{equation} \label{laopacidad}
\kappa_{\nu} \approx  9.78 \times
10^{-3}\frac{n_e}{\nu^2T^{3/2}}\sum_iZ_i^2n_i \times 
\begin{cases} 
18.2 + \ln(T^{3/2})- \ln\nu, & T<2 \times 10^5 K \\ 
24.5 + \ln(T)- \ln\nu, & T > 2\times 10^5 K  \\
\end{cases}
\end{equation}

The source function is:
\begin{equation}\label{leplank}
I_\nu  = \frac{2h\nu^3}{c^2}\frac{1}{\exp{(h\nu/kT)}-1},
\end{equation}
Although, at radio wavelengths $h\nu << kT$, it is possible to use the 
Rayleigh-Jeans
approximation:
\begin{equation}\label{laemision}
I_\nu  \approx \frac{2k\nu^2}{c^2}T,
\end{equation}
Eq. \ref{laopacidad} and \ref{laemision}  are  solved by our model.
We have simulated the solar emission, in the radio wavelengths range, 
and found a good agreement with
observations (Section \ref{sec:results}).

\subsection{The Minimal Emission Parameter}\label{sec:minparameter}
\begin{figure}
\begin{center}
\includegraphics[width=1.0\textwidth]{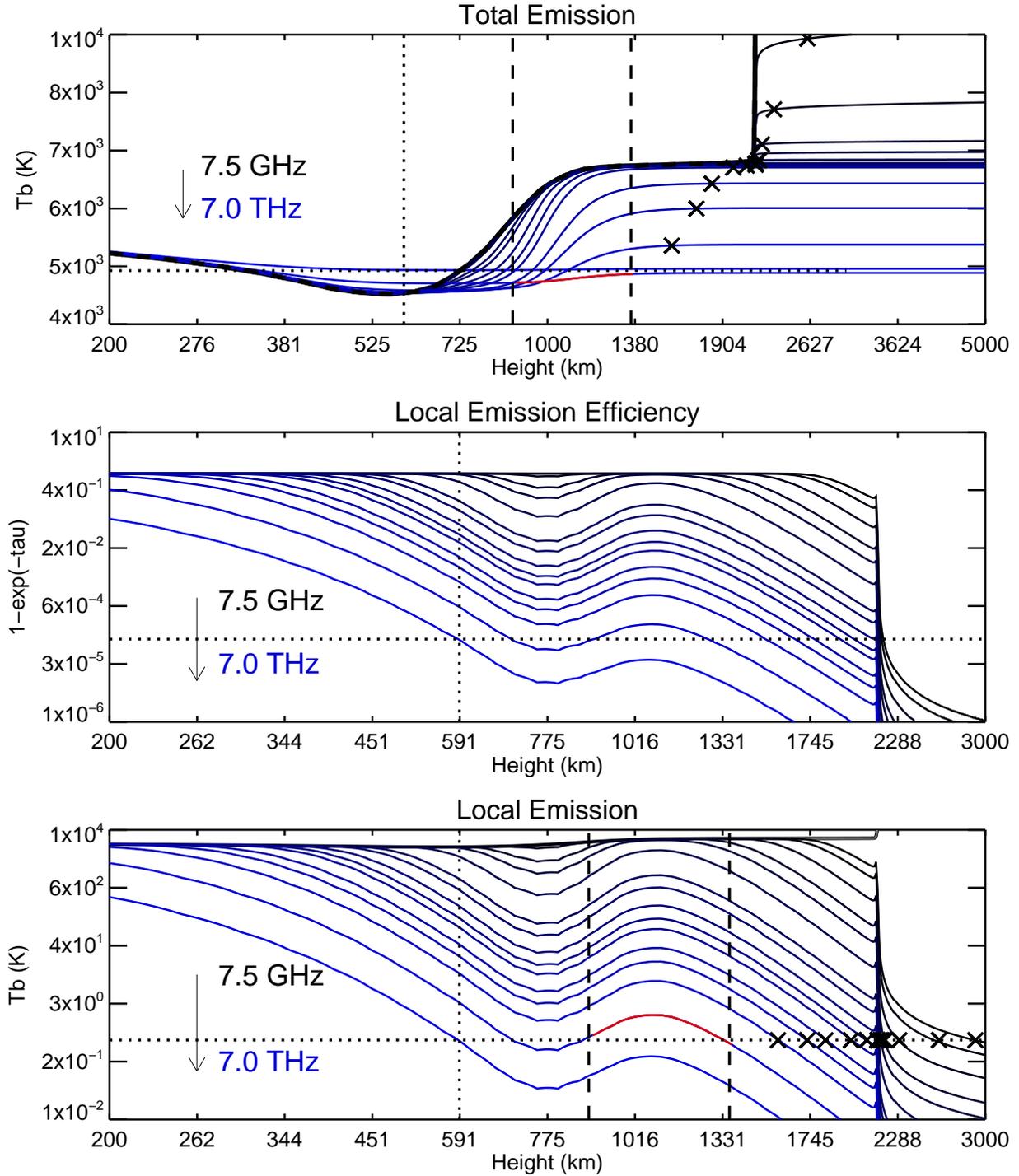}
\caption{Multi-frequency spectrum from 7.5 GHz (black or upper curve) to 7 THz (blue or lower curve).  
The total emission (upper panel), ``emission efficiency'' (middle panel) and
local emission (bottom panel)  as a function of  
height from the photosphere. The thick black line in the upper panel 
is the radial temperature model.
%
}\label{figmin}
\end{center}
\end{figure}

As shown in the upper panel of Figure \ref{figmin}, where we have plotted 
the total emission as a function of the photospheric height for different 
frequencies, from 7 GHz (black curve) to 7 THz (blue curve), above $3000$ km the total emission have reached its final value for all
frequencies. Obviously, this convergence occurs at different heights 
depending on the 
frequency, the minimum height of convergence ($\sim590$ km, marked with a
vertical 
dotted line) corresponds to the 7 THz profile. We use this point as
a reference height ($h_c$). 

On the central panel of Figure \ref{figmin}, we have plotted the 
``emission efficiency'',
$$I_{\mbox{eff}} = 1-\exp(-\tau_{\nu}(z)),$$
 as a
function of height for the same frequency range. Clearly the 7 THz
profile has the  lowest ``emission efficiency''  at all heights. Therefore,
we can use the value of the ``emission efficiency'' of this profile at 
$h_c$, this is, $I_{\mbox{eff}} = 1\times 10^{-4}$, as the lower bound of 
the model (marked with a horizontal dotted line). 
The ``emission efficiency''
can reach lower values, at higher altitudes, but as seen in the upper panel, 
the contribution to the total emission (for the 7 THz profile) at these heights 
is negligible.

As we do not know, before the computations, where the ``emission efficiency''
will reach this lower bound value, we check in the temperature model 
(thick line
in the upper panel) and see that the temperature model reach its minimum 
value $\mbox{MIN}(T_R)$ at $h_c$. Therefore, by using  
Eq. \ref{laemision} we are
able to obtain the minimal significant emission, 
\begin{equation} \label{eq:imin}
I_{\mbox{min}} < \frac{2k\nu^2}{c^2}\mbox{MIN}(T_R)\times10^{-4},
\end{equation}
where $\mbox{MIN}(T_R)$ is the minimum in the 
atmospheric temperature radial profile.

In order to show the correctness of the previous analysis, in the bottom panel
of Figure  \ref{figmin} we have plotted the local emission profiles as a
function of height for the same range of frequencies. The horizontal dotted 
line represents $I_{min}=0.44 K$, computed using  $\mbox{MIN}(T_R)$ at 7 THz. 
As expected, this line intersects with 7 THz profile exactly at $h_c$. 

As an example, we have marked (red line) a not 
negligible excess (i. e. above the dotted horizontal line) of local emission 
at 3 THz, from $\sim800$ to $\sim1300$ km of height (marked with vertical 
dashed lines). And, as shown in the upper panel by a red line in the 3 THz 
profile, only this excess contributes to the total emission.

For lower frequencies we have marked with crosses the height where the local 
emission becomes negligible (this is, where each profile crosses the $I_{min}$ bound
in the bottom panel). This height is also marked with crosses in the
total emission profiles (upper panel), showing that the emission at each 
frequency already have converged to its final value, at the marked height. 

\begin{figure}
\begin{center}
\includegraphics[width=1.0\textwidth]{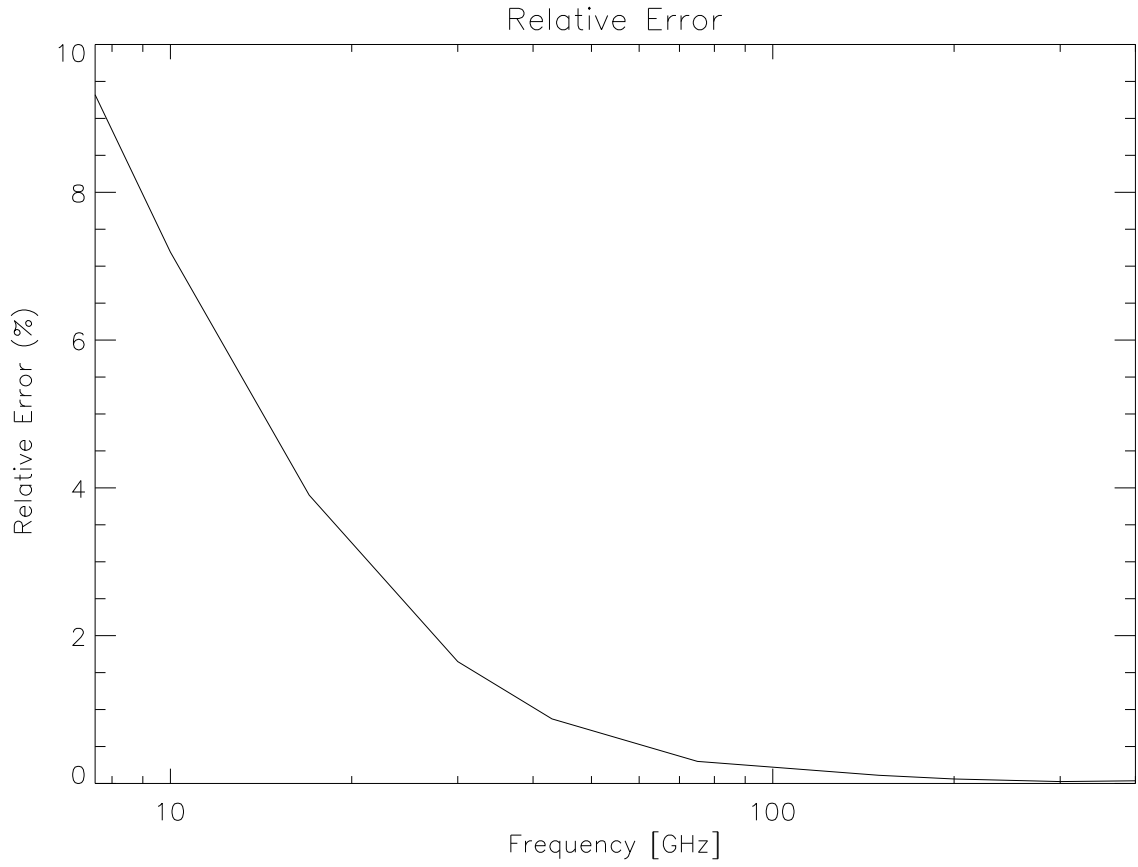} 
\caption{Relative error in the final brightness temperature using  $min = I_{min}$ parameter.}\label{relative_error.ps}
\end{center}
\end{figure}

As shown by Figure \ref{relative_error.ps} where we have plotted the error associated 
with  Eq. \ref{eq:imin}, for 
frequencies higher than $\sim 40$ GHz, 
the relative error of the final brightness temperature is lower than 1\%. 
Whereas for lower frequencies the error is higher, due to the fact 
that there are large regions (at coronal heights) which contribute 
with low amounts of local emission.

\section{Results}\label{sec:results}
\begin{figure}
\begin{center}
\includegraphics[width=1.0\textwidth,]{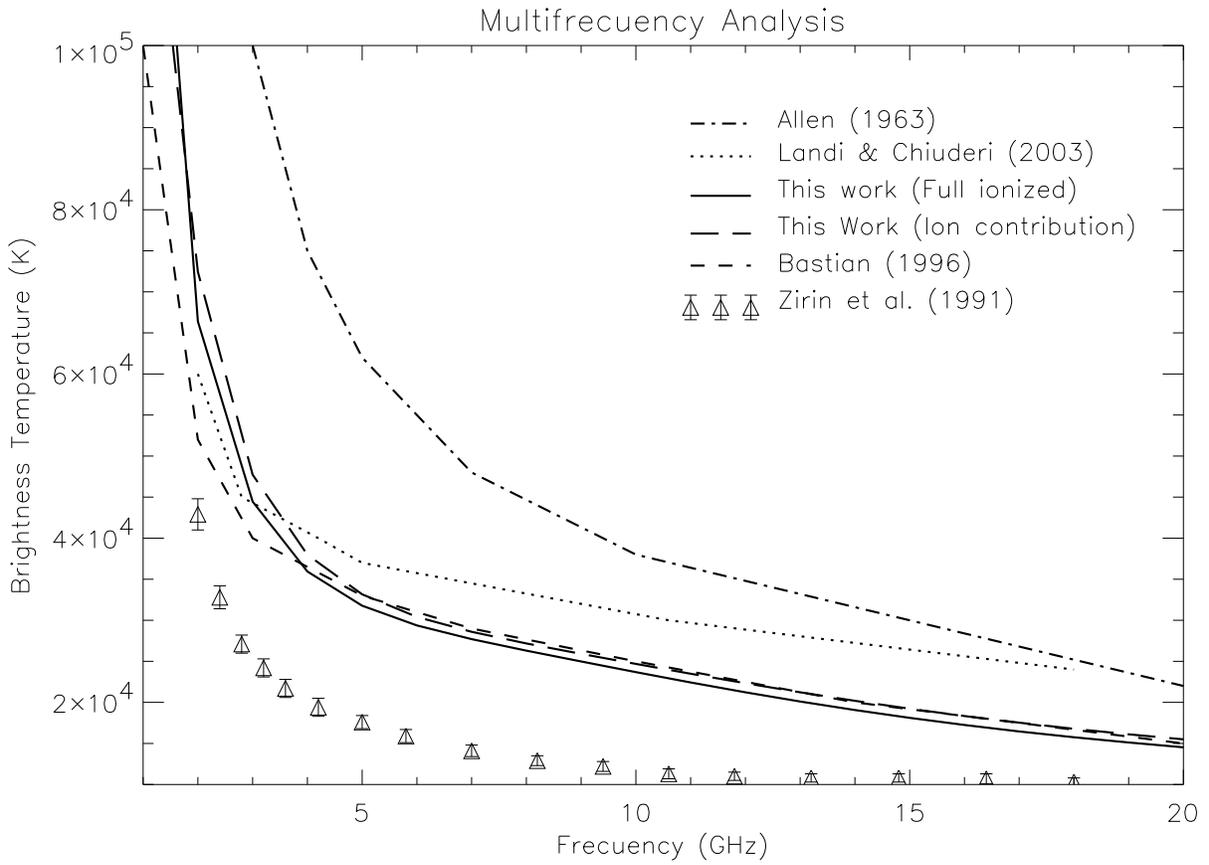} 
\caption{Quiet Sun 
  brightness temperature as a function of  frequency  
from different models. Triangles show the observations of 
\cite{1991ApJ...370..779Z}.}\label{pakalmf}
\end{center}
\end{figure}

We compute the free-free thermal radio emission
from an atmosphere of Hydrogen-Helium
gas, using published (radial) profiles of solar 
temperature and density. 
We performed a multi-frequency analysis, from 2 to 20 GHz, shown in 
Figure \ref{pakalmf} by the continuous and long-dash lines, and compared our
results against observations reported by \cite{1991ApJ...370..779Z} (triangles) and similar
published analysis. 
The continuous line is the output of our model 
using  $n_i=n_e$ in Eq. 
\ref{fullopa}.
The long-dash line
is the  output of our model  considering 
radiation from HII, HeII and HeIII ions in Eq. 
\ref{fullopa}. 
The short-dash line is the \cite{1996ApJ...473..539B} model which use 
similar physical considerations as our model. 
The dotted line
is the 
\cite{2003ApJ...589.1054L} model computed from the observed differential 
emission measure and using an empirical opacity function. 
We also plotted the 
\cite{1963IAUS...16....1A}
(dot-dash line) model.

In Figure \ref{diff},  we have plotted the brightness 
temperature difference, between observations and models, whit the same
line code as Figure \ref{pakalmf}.
This difference decreases with
frequency. At $\sim 5$ GHz, our models and \cite{1996ApJ...473..539B} model
have an excess of $\sim 1.5 \times 10^4$ K, whereas 
\cite{2003ApJ...589.1054L} model has an excess
of  $\sim 2 \times 10^4$ K.
At  $\sim 20$ GHz, all models have better agreement with observations,
although the excess computed by our models as well as \cite{1996ApJ...473..539B} model
is only  $\sim 5 \times 10^3$ K.
 
%

As our code uses a cellular automaton and an expert system to solve 
efficiently the radiative transfer equation, we are able to achieve 
integration times which are up to one 
order of magnitude shorter than direct integration codes (see Appendix \ref{apend:conv}), this makes possible
to
generate high definition 2D images from 3D structures, in reasonably 
short times and using very short (1 km) integration
steps. Therefore, Pakal can compute the emitting spectrum from  
highly detailed source structures, 
as the expected in new generation solar chromospheric models.

We have compared the performance of our code against a similar code published
by \cite{2005A&A...433..365S} and against a linear integration process 
(See Appendix \ref{apend:conv}). We found that
 Pakal can improve the integration time up to one order of magnitude 
compared with the linear integration process and up to 1/3 when compared 
with \cite{2005A&A...433..365S} code.
We have performed a detailed analysis of the quiet Sun emission at 17 GHz
simulated by Pakal and using temperature and 
density profiles observed in UV and continuum 
\citep[see details in][]{2008GeofI..47..197D}.

\begin{figure}
\begin{center}
\includegraphics[width=1.0\textwidth]{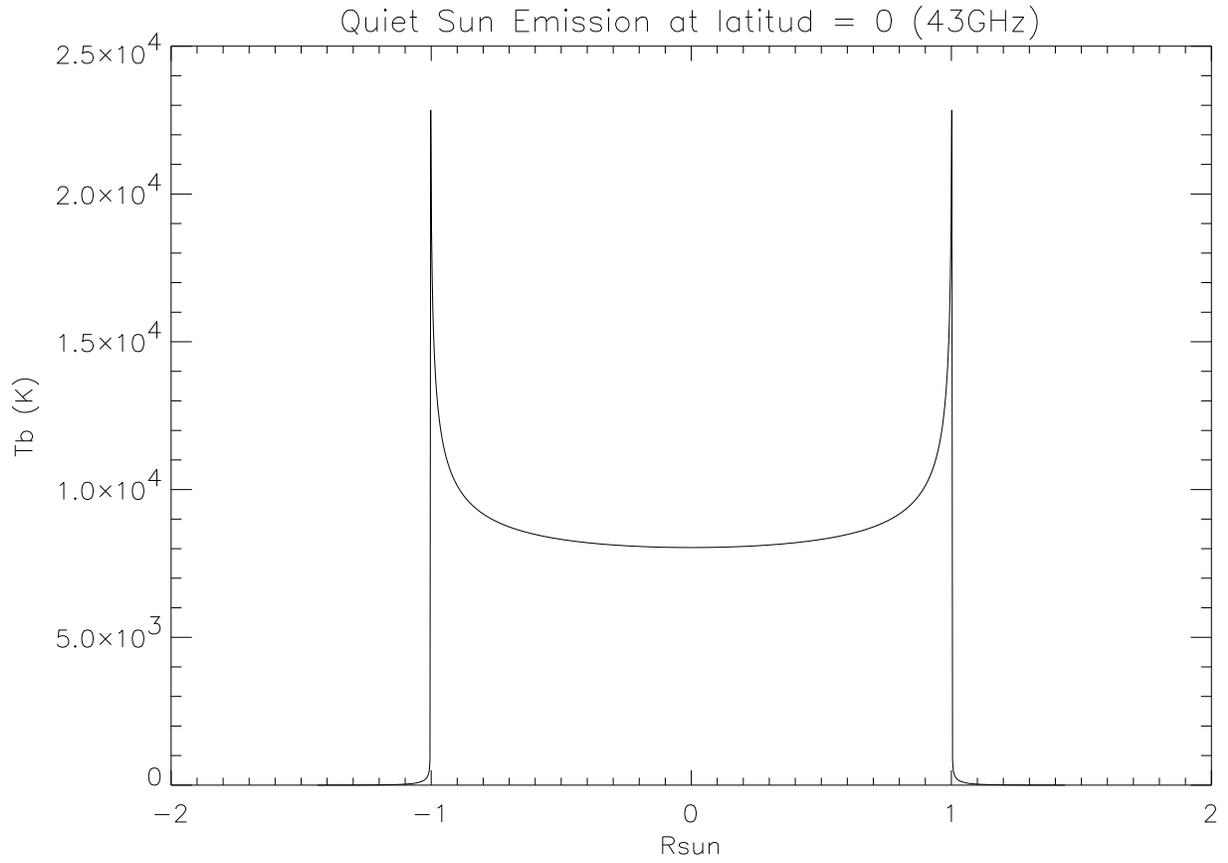}
\caption{A equatorial cut from 1024x1024 simulation image of the Sun at 43 GHz. The figure show the classical limb brightening.}\label{limbo43GHz}
\end{center}
\end{figure}

Figure  \ref{limbo43GHz} shows an equatorial cut of a 1024 by 1024 image 
 of the computed quiet Sun emission at 43 GHz, where the limb brightening is 
clearly seen. 
In this case, we used integration steps of 10 km and a minimal
local emission of $10^{-17}$. We ran the code using the initial values shown in
 Appendix \ref{sec:testing}.
The limb brightening show a maximum intensity of  23000 K and a minimum of 8000 K.
Observations at similar frequencies, made in the 1950's,
reported brightness temperatures from  5700 K to 6000 K at 
40 GHz \citep{1956AJ.....61Q.192W}. Although, based in later observations 
at 50 GHz \citep{1971SoPh...16...75R} predicted a higher emission at the
center of the disk of 7500 K.

\section{Summary}
We have developed a new numerical code to solve the radiative transfer equation
in a radial (3D) geometry for stellar atmospheres.
The code is composed by four independent modules: i) numerical model; 
ii) geometry; iii) numerical methods; and iv) physical stellar models.
This architecture allows  easy changes when we want  to test
different physical models.

We found that the minimum of the
temperature profile, can be used to compute the lower boundary of the emission,
this boundary guarantees the numerical convergence of the final brightness 
temperature.

By improving the geometry and
the integration process, the code is able to reproduce, with better results, 
classical
analysis of the solar radio emission, as the analysis of
the depth of emission 
and
multi-frequency analysis in 1D
; or 2D analysis of 
the limb brightening (Appendix \ref{sec:results}).

The code is up to one order of magnitude faster than linear integration codes,
and three times faster than  similar published codes.
As future work we are going 
to implement adaptative integration steps and
 develop the Message Passing Implementation 
(MPI) of the code which will 
work in multi processors-computers, with these improvements, the code 
will be able to  solve the  
radiative transfer equation in non homogeneous structures with more 
complicated physical conditions as non-LTE, more chemical species and 
emission processes.
Finally, the code is free under request to the author.
\acknowledgments

This work was supported by UNAM-PAPIT grant IN117309 and CONACyT grant 49395
Thanks to  Dr. R. Caballero for allow us to use his computer facilities.

\bibliographystyle{apj}
\bibliography{libros}

\appendix

\section{Testing the model} \label{sec:testing}
\begin{figure}
\begin{center}
\includegraphics[width=1.0\textwidth,]{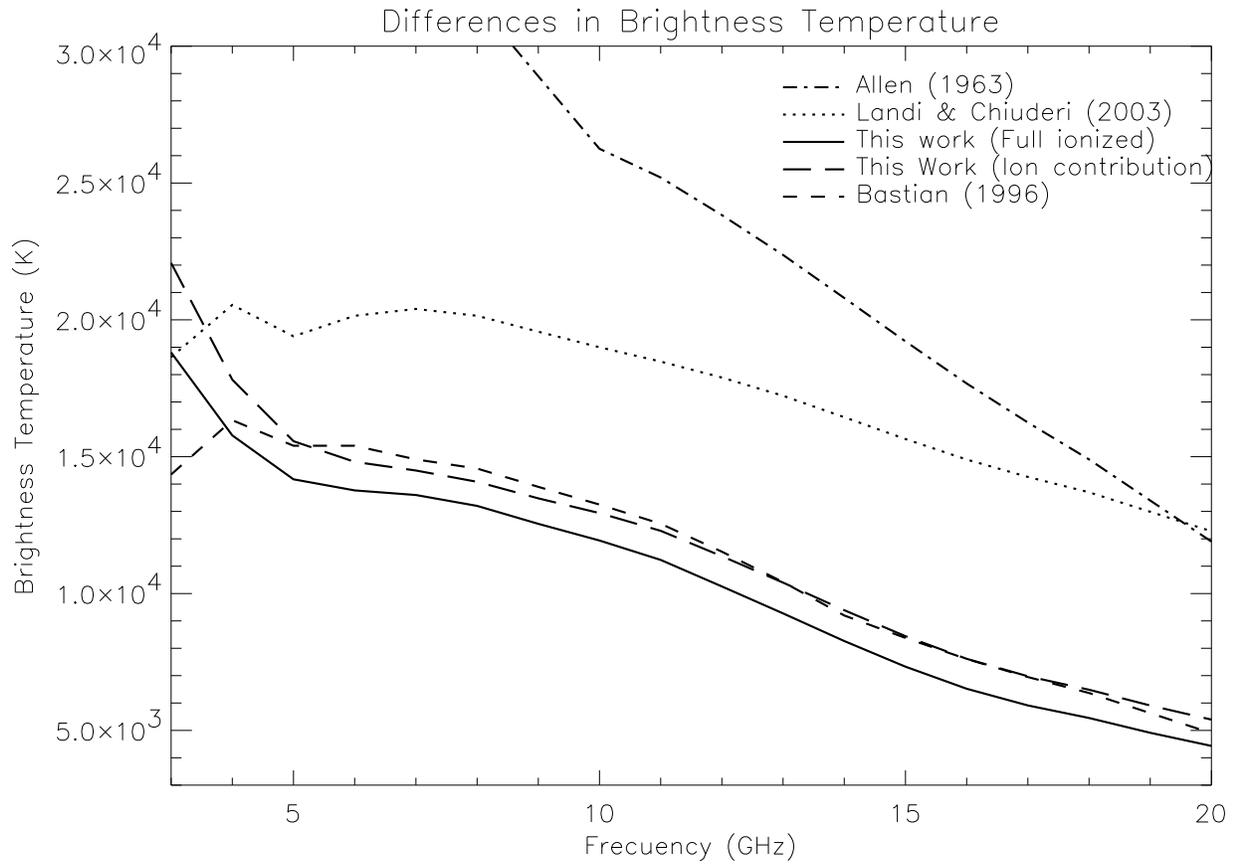}
\caption{Brightness temperature difference between observations and 
models as a function of frequency.}\label{diff}
\end{center}
\end{figure}

Pakal is able to deal with different opacity functions and chemical species 
with different ionization states. Although, in order to
compare our code with previous published results,  
we ran the code twice, firstly, using a restricted initial 
value of densities, $n_i = n_e$   (as shown by continuous lines in 
Figures \ref{pakalmf} and \ref{diff}),
and secondly, considering the density of each ionization state of a diatomic 
gas formed by H-He  
(long-dash lines in Figures \ref{pakalmf} and \ref{diff}).
The code inputs where:
\begin{enumerate}
\item Inputs from libraries:
  \begin{itemize}
  \item Source function: Rayleigh-Jeans approximation (Eq.
    \ref{laemision}).
  \item Opacity: free-free emission 
(Eq. \ref{laopacidad}).
  \end{itemize}
\item Inputs from files:
  \begin{itemize}
  \item Radial profiles of temperature, electronic and Hydrogen densities: 
Here we use the model 
    C of \cite{1981ApJS...45..635V}, for chromospheric and low 
transition zone heights.
For coronal heights,  we use the model
    of  \cite{1976RSPTA.281..339G} and reported by
    \cite{1990soas.book.....F}.
  \item  Assuming  $He = 0.1*H$.
\end{itemize}
\item Console inputs: These inputs changed for each particular 
simulation.
\end{enumerate}

\section{Analysis of Convergence} \label{apend:conv}
The convergence test is necessary when we want 
to prove the adequate 
functionality of any code. 
We have developed three convergence tests, which also help us
to test the efficiency of the code.
The first one involves the -detail parameter, which determines
the length of the integration step when the code is performing a 
detailed integration process. 
The second test involves the -min parameter, which 
sets the minimum emission considered by the code (i. e.,
emission below this value is neglected). 
Finally, we analyze
the -big parameter, which determines the length of the large integration
step 
used
when 
the local emission is negligible. 

In order to found a lower boundary for the minimum emission parameter 
(-min), we ran several simulations at different frequencies. 
We found that the -min parameter have not negligible effects when it 
is greater than the emission computed at  1\% of the minimum of the
temperature profile. Note that the final error of the model is
 associated to this parameter, 
at least 
the total error will be comparable to the minimum emission
parameter
and depends indirectly on the minimum step of integration.

\begin{figure}
\begin{center}
\includegraphics[width=1.0\textwidth]{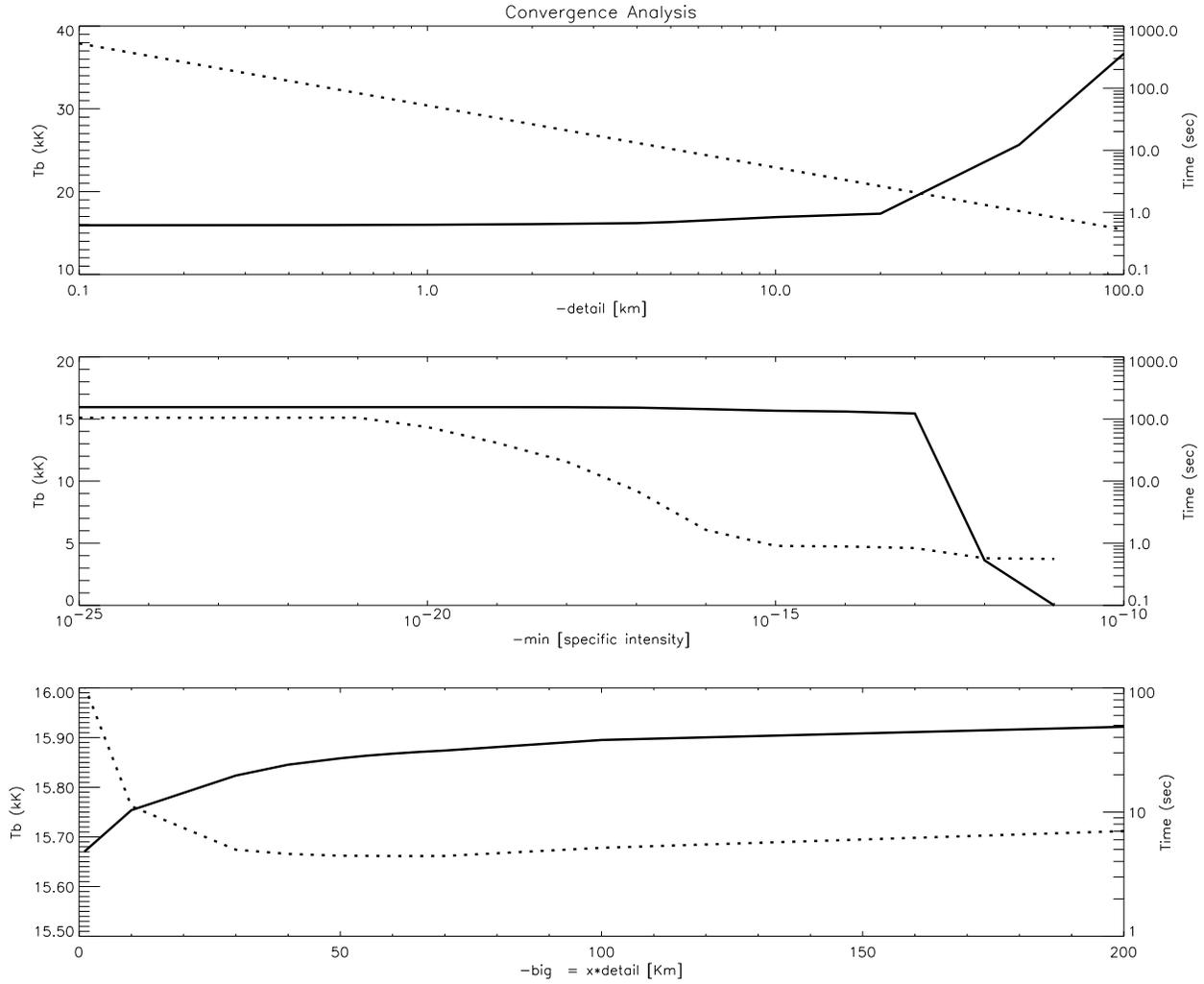}
\caption{Analysis of convergence of Pakal. The brightness temperature 
(continuos lines) and the time of the integration process (dotted lines) 
for the variation of ``-detail'' (upper panel), ``-min'' (middle panel), 
and ``-big'' (bottom panel) parameters.}\label{newConvergence}
\end{center}
\end{figure}

The combination of these 
parameters determine the efficiency of the code.
If we use a very small number for the  -detail parameter, the code will take
long time for the integration process. On the other hand, the code will loose 
valuable information by using too large numbers in the  -big or -min parameters.
Therefore, we need to look for the best parameters in terms of the
integration time and the stability of the output.
In  Figure \ref{newConvergence} 
we have plotted the brightness temperature (continuos lines) and the 
integration time (dotted lines),
versus the varying 
parameter (-detail, -big and -min, respectively) so we can test the 
stability and performance of the code.
The main idea is to find out the best parameters in 
of the shortest integrations times, but without affecting the final 
brightness temperature. 
Those tests were carried out by
computing the emission over a
ray path in a single pixel at position (0,0), i. e., in the center of
the solar disk. 

The upper panel of Figure \ref{newConvergence} 
shows the first test,
the computed brightness temperature as a function of the small 
integration step (-detail parameter) using a constant large integration 
step of 100 km.
If we set the -detail parameter to 100 km, (i. e.,  the detail
integration and the big integration steps are equal), the resultant
algorithm is really poor, because it is integrating sequentially. 
In this case, the
integration time (dotted line in the upper panel of Figure \ref{newConvergence})
is very fast ($\le 1$ sec), but the brightness temperature computed is far away from the
 right value ($1.6 \times 10^{4} K$). 
When the detailed integration steps are lower than 20 km, the emission 
converges to
$1.6 \times 10^{4} K$, 
although the 
integration time grows exponentially. 
As instance, to generate an image of 1024 by
1024 using a small integration step of 10 km, the integration time is almost
two months. If the small integration step is 1 km, the integration 
time will be around two years.

To perform the second test,
 we left constant both: the
small (0.5 km) and large (100 km) integration steps, and 
allow variations of the minimal emission (-min). 
The continuos line in the middle panel of Figure \ref{newConvergence} shows 
that 
the brightness temperature converges when the 
minimal emissions is lower than $10^{-13}$. 
When the minimal emission is higher than $10^{-13}$,
the brightness temperature diverges and the integration times are shorter. As
instance, if the minimal emission is $10^{-17}$ and the integration step 
of 0.5 km, an image of 1024 by 1024 takes 85 days of integration.
 
In order to find out the best value for both integration steps (third test), 
in the buttom panel of Figure \ref{newConvergence}  we have plotted the 
brightness temperature (continuos line) as a function of the
large integration step ,
in terms of the small integration step ($large = n \times small$), 
setting the
minimal emission as $10^{-17}$ and the small integration
step as 0.5 km.

\begin{figure}
\begin{center}
\includegraphics[width=1.0\textwidth]{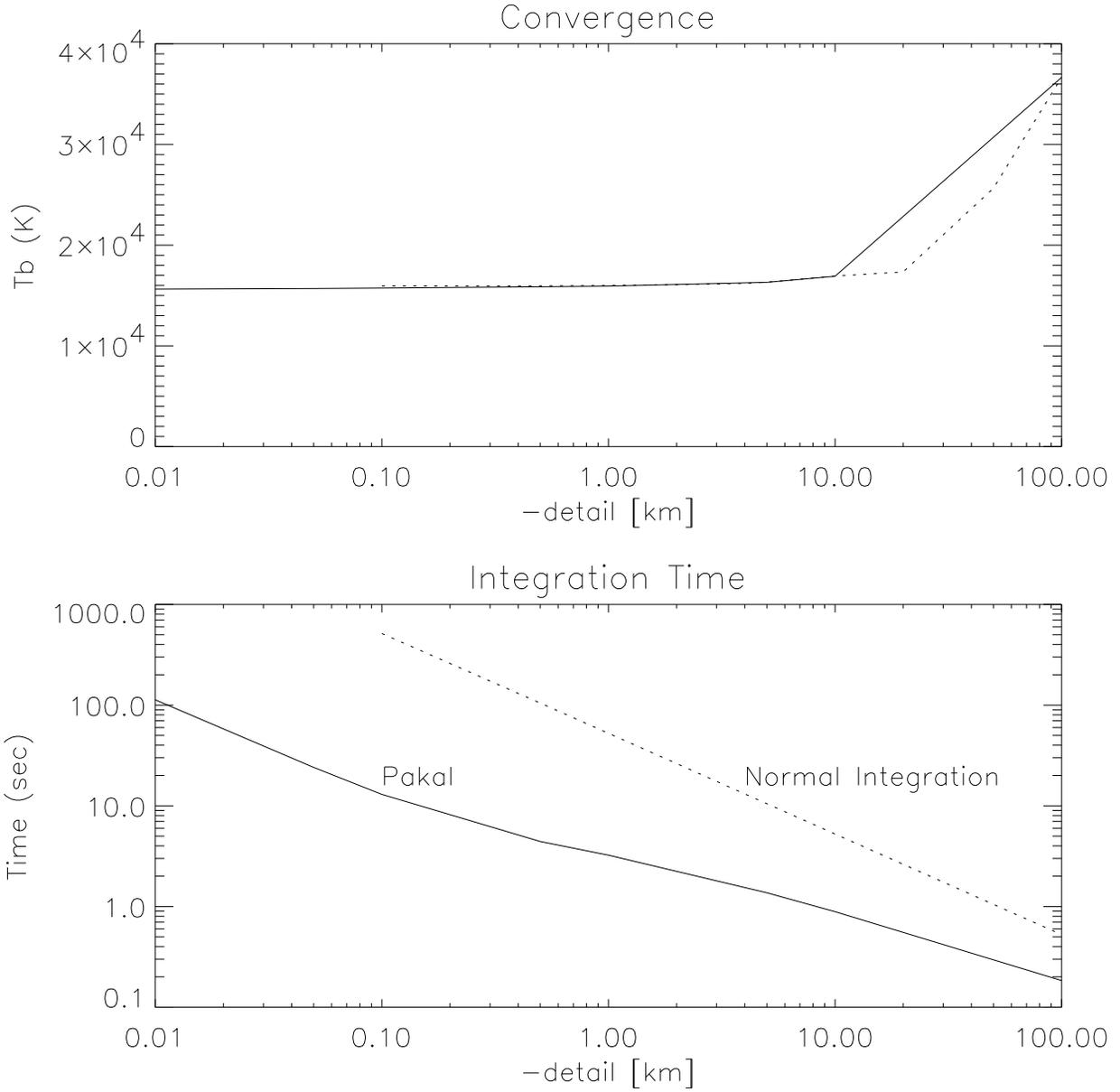}
\caption{Analysis of convergence of Pakal. 
The brightness temperature as function of the short integration steps and
using the optimal parameters; calling sequence:
``./pakal -xy 0 0 -nu 17e9 -min 1e-17 -detail X -big 60''.
The bottom panel shows the respective time (continuous line)and the time using
linear integration (dotted line).
}\label{convergenciasuper}
\end{center}
\end{figure}

Changes of the large integration step do not affect appreciably
the brightness temperature, although, the integration time is largely 
affected by such changes. In this case,
the integration time may vary in one order
of magnitude. The minimum time of integration  is reached at 60 km (see 
the dotted line of the button panel of Figure \ref{newConvergence}),
this is:
$$big[\mbox{km}] = 60*\mbox{detail} = 30 km.$$
Performing the convergence analysis, but using the best parameters,
is possible to obtain integration times which are one order of magnitude
lower than direct integration process, 
as shown in Figure \ref{convergenciasuper} 
For instance, the integration time for a 1024 by 1024
image with small integration steps of 10 km is now 11 days instead of 
two months.
If the small integration step is 1 km Pakal now takes 39 days instead of 
two years
(a super computer with 1024 processors will take one hour to generate this 
image).

We have compared the performance of our code against a similar code published
by \cite{2005A&A...433..365S} which practically solves the same task but 
based on linear integration method. The border conditions are:
\begin{itemize}
\item Spatial resolution: 1100 points = 770000 km.
\item Steps of Integration: 800.
\item Size of a step: 50 km.
\end{itemize}
The time results are as follows:
\begin{itemize}
\item Selhorst code: ~13 minutes.
\item Pakal code: ~33 minutes.
\end{itemize}
Changing the Pakal integration parameters to the optimal value, 
\begin{itemize}
\item Detail integration step: 50 km. 
\item Large integration step: 100*50 km.
\item Minimal emission parameter: 1e-20.
\end{itemize}
We obtain 
\begin{itemize}
\item Pakal code: ~4 minutes.
\end{itemize}
Summarizing we improved in one order of magnitude
our results and in 1/3 the linear integration used in another method.
Even more, that the method used by \cite{2005A&A...433..365S}
 does not include the computation of the geometry of the problem.



 \begin{table}[h]
\begin{center}
  \begin{tabular}{| c | c | c | l | c | c |}
\hline                 
   State & q & y & Instructions & $\epsilon$ & Step (dz)\\
\hline
A1 & 0 & 0 & $I = 0$          & 0 & small \\
   &   &   & $x_a = x_b$        &   &  \\
   &   &   & $x_a += dzdetail$ &   &\\
\hline
   & 1 & 1 & $x_a = x_b$        & 1 & large  \\
   &   &   & $x_a += dzbig$    &   & \\
\hline
   & 2 & 0 & $I = I_oe^{-\tau} + S(1-e^{-\tau})$ & 0 & small\\
   &   &   & $x_a = x_b$        &   & \\
   &   &   & $x_b += dzdetail$ &   & \\
\hline
A2 & 0 & 0 & $I = 0$          & 0 & small \\
   &   &   & $x_a = x_b$        &   &  \\
   &   &   & $x_b += dzdetail$ &   &\\
\hline
   & 1 & 1 & $x_a = x_b$        & 1 & large  \\
   &   &   & $x_b += dzbig$    &   & \\
\hline
   & 2 & 0 & $x_b = x_a+dzdetail$  & 2 & small\\
\hline
A3 & 0 & 0 & $I = 0$          & 0 & small \\
   &   &   & $x_a = x_b$        &   &  \\
   &   &   & $x_b += dzdetail$ &   &\\
\hline
   & 1 & 1 &$I = I_oe^{-\tau} + S(1-e^{-\tau})$  & 1 & small  \\
   &   &   &  $x_a = x_b$         &   & \\
   &   &   & $x_b += dzdetail$    &   &\\
\hline
   & 2 & 0 & $I = I_oe^{-\tau} + S(1-e^{-\tau})$ & 1 & small\\
   &   &   & $x_a = x_b$        &   & \\
   &   &   & $x_b += dzdetail$ &   & \\
\hline

  \end{tabular}
  \caption{Decision Table of the Coordinator. Where $I$ is the local emission after the computation; 
$I_o$ is the incoming emission; $S$ is the source function; $\tau$ the local opacity; $x_a$ and $x_b$ the two spatial coordinates; $dzdetail$ the small integration step; and $dzbig$ the large integration step. Using this decision table the coordinator chooses the integration step and computes the local emission $I$.}\label{tabladecision}
\end{center}
\end{table}

\begin{table}[h]
\begin{center}
  \begin{tabular}{| c | c | c |}
\hline                 
   Variable & Value & Meaning\\
\hline
$y$ & 0 &  Next step small.\\
  & 1 &  Next step large.\\
  & 2 &  Next step backwards\\
\hline
$q$ & 0 & The wave can not propagate.\\
  & 1 & The emission is not enough.\\
  & 2 & There is enough emission.\\
\hline  
  \end{tabular}
  \caption{State of the expert system,
where  $q$ and $y$ are the possible recommendations. 
 }\label{tablayq}
\end{center}
\end{table}

\begin{table}[h]
\begin{center}
  \begin{tabular}{| c | c | c | c |}
\hline                 
   State & $\epsilon$ & New State & Execute \\
\hline
A1 & 0 & A1 & Nothing \\
   & 1 & A2 & Nothing \\
   & 2 & A4 & Nothing\\
\hline
A2 & 0 & A1 & Nothing \\
   & 1 & A2 & Nothing \\
   & 2 & A3 & $i = 1$ \\
\hline
A3 & 0 & A1 & Nothing \\
   & 1 AND $(i < n)$ & A3 & i++ \\
   & 1 AND $(i = n)$  & A1 & Nothing \\
   & 2            & A4& Nothing\\
\hline
A4 & 0 & null & Error \\
   & 1 & null & Error \\
   & 2 & null & Error \\
\hline
  \end{tabular}
  \caption{Table of States of the Cellular Automaton. In this case $n=dzbig/dzdetail$. The variable ``state'' represents the memory stages of the integration process; 
``$\epsilon$'' controls the switch between states; and the variable ``i'' is a stack into the automaton.}\label{tablaestado}
\end{center}
\end{table}

\end{document}